\documentclass[printer]{aa} % for a referee version
\usepackage{relsize}

\newcommand{\out}[1]{}  %%take out (hide) text within command domain

\newcommand{\ngc}{\hbox{NGC\,2363}}
\newcommand{\pap}{\hbox{Paper\,I}}
\newcommand{\dri}{\hbox{DUCBR}}

\newcommand{\msol}{\hbox{${\rm M_{\sun}}$}}

\newcommand{\cms}{\hbox{${\rm cm^{-2}}$}}

\newcommand{\gcc}{\hbox{${\rm g\,cm^{-3}}$}}
\newcommand{\cc}{\hbox{${\rm cm^{-3}}$}}
\newcommand{\cmc}{\hbox{${\rm cm^{-3}}$}}
\newcommand{\kms}{\hbox{${\rm km\,s^{-1}}$}}

\newcommand{\flu}{\hbox{${\rm erg}\,{\rm cm}^{-2}\,{\rm s}^{-1}$}}
\newcommand{\fvun}{\hbox{${\rm erg}\,{\rm cm}^{-2}\,{\rm km}^{-1}$}}
\newcommand{\fv}{\hbox{$F^{v}_{\rm 15}$}}

\newcommand{\ergs}{\hbox{${\rm erg\, s^{-1}}$}}

\newcommand{\thei}{\hbox{$\theta_{i}$}}
\newcommand{\thef}{\hbox{$\theta_{f}$}}

\newcommand{\cl}{\hbox{$\overline{s}$}}
\newcommand{\rhol}{\hbox{$\overline{\rho}$}}

\newcommand{\vw}{\hbox{$V_{w}$}}
\newcommand{\vli}{\hbox{$V_{lim}$}}

\newcommand{\nw}{\hbox{$n_{w}$}}

\newcommand{\no}{\hbox{$n_{o}$}}

\newcommand{\Tw}{\hbox{$T_{w}$}}
\newcommand{\Tx}{\hbox{$T_{X}$}}

\newcommand{\To}{\hbox{$T_{o}$}}

\newcommand{\Po}{\hbox{$P_{o}$}}

\newcommand{\gam}{\hbox{$\Gamma$}}
\newcommand{\ep}{\hbox{$\epsilon$}}
\newcommand{\lo}{\hbox{$l_{o}$}}

\newcommand{\Uo}{\hbox{$U_o$}}

\newcommand{\phiw}{\hbox{$\varphi_{w}$}}
\newcommand{\phix}{\hbox{$\varphi_{X}$}}
\newcommand{\phis}{\hbox{$\varphi_{\ast}$}}

\newcommand{\Nx}{\hbox{$N_{X}$}}
\newcommand{\No}{\hbox{$N_{\rm o}$}}
\newcommand{\Ntw}{\hbox{$N_{\rm 20}^{\rm o}$}}

\newcommand{\map}{\hbox{{\sc mappings i}c}}
\newcommand{\fwhm}{\hbox{{\sc fwhm}}}
\newcommand{\fwzi}{\hbox{{\sc fwzi}}}
\newcommand{\sed}{\hbox{{\sc sed}}}

\newcommand{\chb}{\hbox{$C({\rm H}\beta$})}

\newcommand{\hb}{H$\beta$}

\newcommand{\ha}{\hbox{H$\alpha$}}

\newcommand{\heii}{\hbox{He\,{\sc ii}}}

\newcommand{\heiiw}{\hbox{He\,{\sc ii}\,$\lambda $4686}}

\newcommand{\neiii}{\hbox{[Ne\,{\sc iii}]}}

\newcommand{\sii}{\hbox{[S\,{\sc ii}]}}

\newcommand{\nii}{\hbox{[N\,{\sc ii}]}}
\newcommand{\niiw}{\hbox{[N\,{\sc ii}]$\lambda $6583}}

\newcommand{\ciii}{\hbox{C\,{\sc iii}]}}

\newcommand{\civ}{\hbox{C\,{\sc iv}}}

\newcommand{\ariv}{\hbox{[Ar\,{\sc iv}]}}

\newcommand{\mgii}{\hbox{Mg\,{\sc ii}}}
\newcommand{\niii}{\hbox{N\,{\sc iii}]}}

\newcommand{\npp}{\hbox{N$^{+2}$}}
\newcommand{\opp}{\hbox{O$^{+2}$}}
\newcommand{\oiii}{\hbox{[O\,{\sc iii}]}}
\newcommand{\oiiiw}{\hbox{[O\,{\sc iii}]$\lambda $5007}}

\newcommand{\oii}{\hbox{[O\,{\sc ii}]}}

\newcommand{\hii}{\hbox{H\,{\sc ii}}}

\usepackage{graphics}
\begin{document}

%   \thesaurus{08
 %             (09.08.2;  % ISM: Herbig-Haro Objects
 %              08.13.2;  % Stars: Mass-loss
 %              09.10.1;  % ISM: Jets and outflows
 %              02.19.1;  % Shock Waves
  %             08.06.2;  % Stars: formation
  %             02.08.1)} % Hydrodynamics
%
   \title{The broad \ha, \oiii\ line wings in stellar supercluster\,A of NGC\,2363
   and the turbulent mixing layer hypothesis}

    \titlerunning{Broad line wings in nebula NGC\,2363}

%   \subtitle{}
\author{Luc Binette
          \inst{1,2}
\and Laurent Drissen
          \inst{1}
\and Leonardo \'Ubeda \inst{1} \and Alejandro C. Raga\inst{3}
          \and \\ Carmelle Robert\inst{1}
          \and Yair Krongold\inst{2}
          }

   %\offprints{L. Binette}

   \institute{D\'{e}partement de physique, de g\'{e}nie physique et d'optique \& Centre
   de recherche en astrophysique du Qu\'ebec,
Universit\'{e} Laval, Qu\'{e}bec, Qc, G1V\,0A6 \and  Instituto de
Astronom\'\i a, Universidad Nacional Aut\'{o}noma de M\'{e}xico, Ap. 70-264,
04510 M\'exico, D.F., M\'exico \and Instituto de Ciencias Nucleares,
Universidad Nacional Aut\'{o}noma de M\'{e}xico, Ap. 70-543, 04510 M\'exico,
D.F., M\'{e}xico
%email: lbinette@astroscu.unam.mx
             }

   \date{}

\abstract{Supercluster\,A in the extragalactic \hii\ region \ngc\ is
remarkable for the hypersonic gas seen as faint extended broad
emission lines with a full-width zero intensity of 7000\,\kms.}{We
explore the possibility that the observed broad profiles are the
result of the interaction of a high-velocity cluster wind with dense
photoionized clumps.}{The geometry considered is that of near static
photoionized condensations at the surface of which turbulent mixing
layers arise as a result of the interaction with the hot wind. The
approximative treatment of turbulence was carried out using the
mixing length approach of Cant\'o\& Raga. The code \map\ was used to
derive the mean quantities describing the flow and to compute the
line emissivities within the turbulent layers. The velocity
projection in three dimensions of the line sources was carried out
analytically.}{A fast entraining wind of up to $\approx 4300\,$\kms\
appears to be required to reproduce the faint wings of the broad
\ha\ and \oiii\ profiles. A slower wind of 3500\,\kms, however, can
still reproduce the bulk of the broad component and does provide a
better fit than an ad\,hoc Gaussian profile.}
%In order to fully cover the full extent of the profile width, the thickness
%of the mixing layers cannot be single valued.
{Radial acceleration in 3D (away from supercluster\,A) of the
emission gas provides a reasonable first-order fit to the broad line
component. No broad component is predicted for the \nii\ and \sii\
lines, as observed. The wind velocity required is uncomfortably high
and alternative processes that would provide comparable constant
acceleration of the emission gas up to 4000\,\kms\ might have to be
considered.}

\maketitle

%
%________________________________________________________________

\section{Introduction}

NGC\,2363, the largest and most massive \hii\ region in the dwarf
galaxy NGC\,2366 (distance = 3.42\,Mpc; Thuan \& Izotov 2005)
harbors one of the best documented cases of hypersonic gas seen as
faint extended broad emission lines in an increasing number of giant
extragalactic \hii\ regions (Tenorio-Tagle et\,al. 1997;
Westmoquette et\,al. 2007a,b,c, 2008). In NGC 2363, this broad
component, first reported by Roy et\,al. (1992), appears as a faint
pedestal under the narrow \ha, \hb\ and \oiii\ lines of the
so-called knot\,A (Gonzalez-Delgado et\,al. 1994). This pedestal has
an \fwhm\ in excess of 2300\,\kms\ (Drissen et\,al. 2009; hereafter
\dri).
%\footnote{See nebula labeled\,I in Fig.\,1 of Drissen et\,al.
%(2000).} over 100\,pc

NGC\,2363 is ionized by two massive star clusters labeled\,A (age
less than 1\,Myr) and \,B (age 3--4\,Myr old\footnote{Knot\,B,
although weaker, contains four Wolf-Rayet stars and one Luminous
Blue Variable (Drissen et\,al. 2001).}; see Drissen et\,al. 2000).
The most intense nebular flux is associated with supercluster\,A,
both in terms of ``normal'' emission lines and of the broad emission
component. The broad component is the focus of this paper.

The origin of the faint broad emission remains unclear.
Tenorio-Tagle et\,al. (1997) proposed that it is caused by  the
breakout of fast expanding shells due to Rayleigh-Taylor
instabilities, which would be significantly delayed in low
metallicity gas and in the presence of a very energetic source.
Westmoquette et\,al. (2007a,b,c) observed faint broad wings in a
number of starburst galaxies and suggested that turbulent mixing
layers (hereafter TMLs) on the surface of gas clumps, set up by the
impact of the fast-flowing cluster winds (Pittard et\,al. 2005),
might account for this phenomenon.

In this paper, we quantitatively explore whether simple TML models
can reproduce the profile shape of the faint line wings observed in
\ngc. In order to test our model, we have used spectra of the
western region close to the center of stellar supercluster\,A.
%using the Gemini Multi-Object Spectrograph (GMOS-N) with the Integral
%Field Unit (IFU). In this paper,
We present succinctly the observations of the broad faint wings in
\ha\ and \oiii\ and compare them with simple TML models. A thorough
analysis of the data set is presented in \dri.

%The paper is organized as follows. The equations describing a
%turbulent mixing layer and their implementation in the code \map\
%are discussed in \S\,\ref{sec:eqn}. The results obtained from
%numerical integrations of the energy and ionization rate equations
%are described in Section~3. The line ratios predicted from these
%models are discussed in Section~4,...

\section{The observations}\label{sec:obs}

In order to test our model, we have used spectra of the western
region close to the center of supercluster\,A described by \dri,
which we briefly describe here. These data were obtained with the
Gemini Multi-Object Spectrograph's Integral Field Unit (GMOS-IFU)
attached to the Gemini North telescope. The two-slit mode was used,
covering a $5\arcsec\times 7\arcsec$ field of view, centered on the
brightest region of nebular emission. Two gratings were used: R831,
covering the 6025--6760\,\AA\ wavelength range with a resolution of
1.4\,\AA, and  B600, covering the 4090--5400\,\AA\ wavelength range
with a resolution of 2.7\,\AA.

The IFU spectroscopic mode allows mapping of the spatial extent and
geometry of the broad component. Using the current data set, \dri\
showed that the intensity of the weak broad component in \ha\ is
2--3\% that of the narrow component, in agreement with the results
of Gonzalez-Delgado et\,al. (1994).

\section{Modeling turbulent mixing layers} \label{sec:eqn}

%Turbulent mixing layers (TMLs) induced by a hot wind is
% Slavin, Shull \& Begelman (1993).

\begin{figure}
\resizebox{\hsize}{!}{\includegraphics{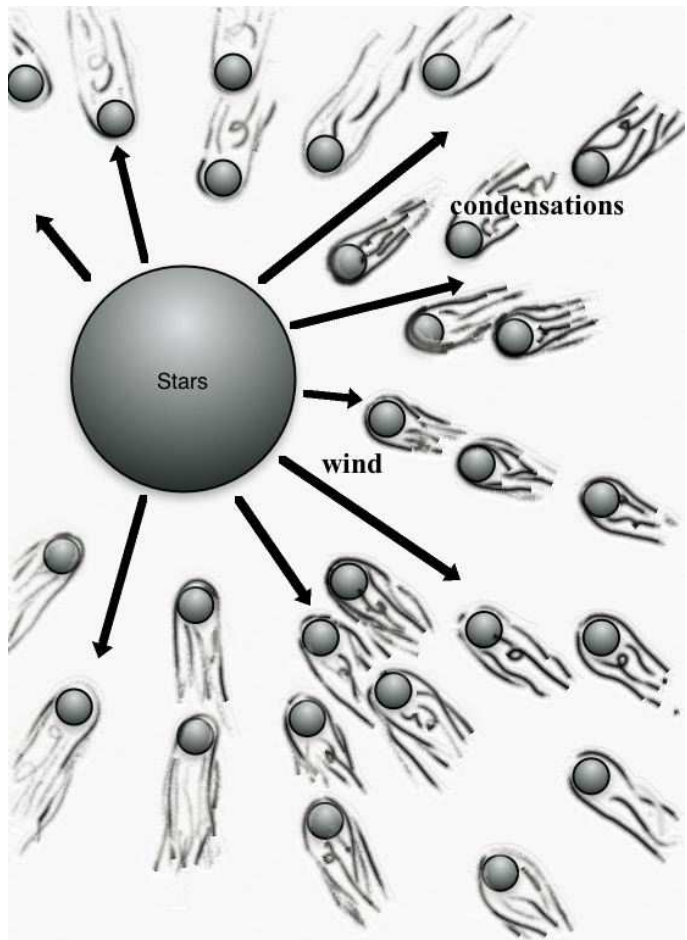}} \caption{The
drawing describing the geometry of the supercluster hot wind within
which photoionized condensations are distributed.
%The TMLs arise from the surface of the gas condensations immersed in the wind.
}
\label{fig:cart}
\end{figure}

In the broader context of the interstellar and the intracluster
medium, Begelman \& Fabian (1990) derived a prescription for
evaluating the temperature of a turbulent mixing layer, which Slavin
et\,al. (1993) used  to calculate the emission line spectrum of such
a layer. Rand (1998) established an interesting comparison between
their model predictions and the spectrum produced by the diffuse gas
observed in the spiral NGC\,891. The models that best reproduce the
observations required a mixture of TMLs and matter-bounded
photoionized condensations.
%Turbulent mixing layers have also been considered for explaining the
%spectrum from low excitation knots observed in Herbig-Haro objects (Binette et\,al. 1999).
More recently, Binette et\,al. (2008; hereafter \pap) improved
earlier models in two ways: a) the hot mixing layer and the warm
photoionized gas are integrated into a single albeit stratified
component exposed to an external ionizing source, and b) rather than
considering a single temperature for the TML given by the
geometrical mean of the warm and hot phases, a temperature structure
is derived using the mixing length scheme presented in the work of
Cant\'o \& Raga (1991). The possibility of computing not only
emission line intensities (Binette et\,al. 1999), but also line
profiles, has now been implemented to allow a comparison of TMLs
with the very broad but faint \oiii\ and \ha\ wings observed in
\ngc.

\subsection{A mixing length approach}\label{sec:app}

\begin{figure*}
\resizebox{\hsize}{!}{\includegraphics{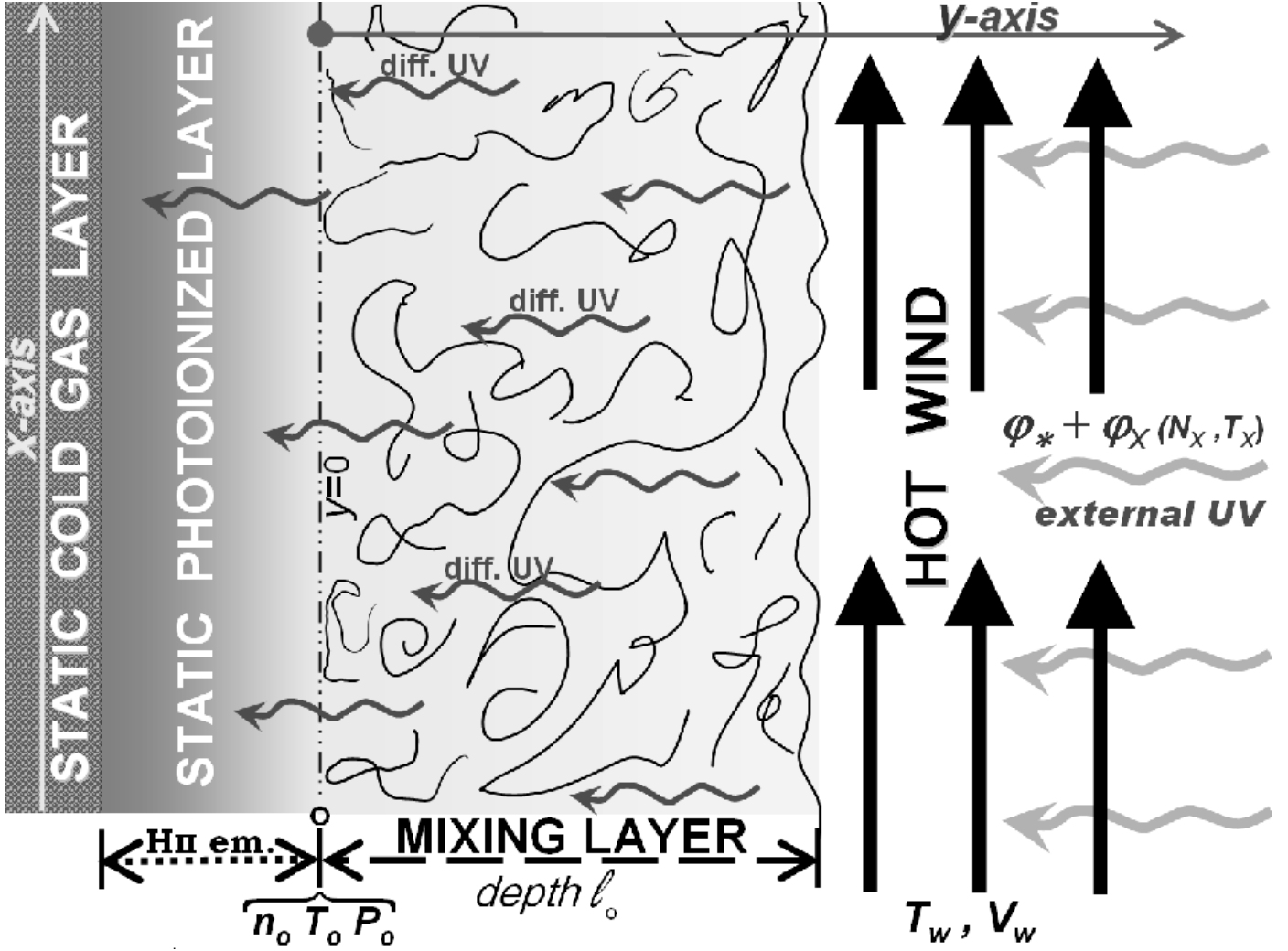}}
\caption{Schematic diagram showing the cross-section along the
y-axis of a plane-parallel mixing layer of thickness \lo.
%The x-axis (parallel to the flow) is structure-less in our equations.
The mixing layer is formed by the interaction of a supersonic wind
(velocity \vw\ and temperature \Tw) with a static layer of warm gas
of H\,number density \no\ and temperature \To\ (at $y=0$).  }
\label{fig:dia}
\end{figure*}

We postulate that both the narrow and the wide emission line
components of the observed profiles arise from ionized gas
condensations of low volume filling factor that are moving randomly.
In a typical extragalactic \hii\ region with an \hb\ luminosity of
$10^{40}$\ergs, this virial turbulence would give rise to a velocity
dispersion\footnote{It is a well documented fact that the line
profiles of giant \hii\ regions greatly exceed the width given by
thermal broadening alone (c.f. Melnick et\,al. 1988, 2000 and
references therein).} of $\sim 25\,\kms$ (Melnick et\,al. 2000). In
\ngc, we postulate the existence of a fast hot wind originating from
the ionizing cluster, which flows around these condensations, giving
rise to a mixing layer at their surface, as depicted in
Fig.\,\ref{fig:cart}. It is this interaction between the fast wind
and the condensations that would give rise to the observed broad
component. To describe the mixing layer, we assume for simplicity an
infinite plane interface, along which a hot wind of temperature \Tw\
is flowing supersonically with velocity \vw\ with respect to a
static warm gas layer of hydrogen number density \no\ and
temperature \To. The turbulent mixing layer that develops between
the hot and warm phases has a geometrical thickness \lo, as
described in the diagram of Fig.\,\ref{fig:dia}. The TML structure,
the static photoionized gas and the hot wind are all isobaric, with
a pressure \Po\ ($\approx 2 k \, \no \To$). The nebular optical
emission lines take their origin within the warm section of the
mixing layer as well as within the static photoionized layers.

Within the TML layer, between $y=0$ and $y=\lo$, the gas is
entrained and accelerated. From the theoretical point of view, the
problem of entrainment in a mixing layer involves the description of
a turbulent flow. However, following the work of Hartquist et\,al.
(1986), Cant\'o \& Raga (1991) developed  an approximate treatment
of this problem based on a mixing length approach of `turbulent
viscosity' and on laboratory measurements. Cant\'o \& Raga (1991)
proposed a "single parcel" model, considering mean values for the
flow variables (density, temperature, velocity) averaged across the
width of the mixing layer. Noriega-Crespo et\,al. (1996) proposed an
alternative approach, developing a model which resolves the cross
section of the mixing layer (in this model, the variables correspond
to averages along the $x$-axis, aligned with the direction of the
shear flow). This latter approach was explored further by Binette
et\,al. (1999). The adoption of such `mean' flow characteristics
allows us to get around having to deal with the details of the
turbulence cascade.
%The mixing layer occurs on the surface of a static photoionized
%layer (labeled {\it \hii\ em.} in Fig.\,\ref{fig:dia}).

For the case of a thin, steady state, high Mach number radiative
mixing layer, the advective terms along the direction of the mean
flow can be neglected with respect to the corresponding terms across
the thickness of the mixing layer (Noriega-Crespo et\,al. 1996,
Binette et\,al. 1999).  Under this approximation, the momentum and
energy equations can be written as:
\begin{eqnarray}
\mu {d^2 v\over dy^2}=0 \label{eq:vel} \; ,\\
\kappa {d^2T\over dy^2}+\mu \left({dv\over dy}\right)^2=L-G \; ,
\label{eq:flo}
\end{eqnarray}
where $y$ is a coordinate measured from the onset of the mixing
layer (see Fig.~\ref{fig:dia}), $v$ the bulk flow velocity (the
velocity component projected along the $x$ axis) as a function of
$y$ , $L$ and $G$ are the radiative energy loss and gain per unit
volume, respectively, and $\mu$ and $\kappa$ are the turbulent
viscosity and conductivity, respectively, which are assumed to be
constant throughout the cross-section of the mixing layer.
Eqn.\,\ref{eq:vel} can be integrated to obtain the linear Couette
flow solution:
\begin{eqnarray}
v(y)={y\over \lo} \, \vw \; , \label{eq:vy}
\end{eqnarray}
where \lo\ is the thickness of the TML and \vw\ is the velocity of
the hot wind, parallel to the static layer. As the gas enters the
TML and starts flowing, its `mean' velocity in the $x$ direction
increases linearly until it reaches the wind velocity at $y=\lo$.
This solution can be substituted in Eqn.\,\ref{eq:flo}, which can
then be integrated to derive the temperature behavior across the
layer.
%Raga \& Cant\'o (1997) integrated this equation analytically with an
%idealized energy loss term. The resulting solutions demonstrate that
%narrow mixing layers are adiabatic (showing a parabolic temperature
%cross-section), and wide mixing layers are radiative (showing a
%flat-topped, quasi-isothermal temperature cross-section).

Within the TML, we consider the equations governing the
\emph{fractional} abundance $f_i$ of each species $i$. This
abundance must satisfy the equation:
\begin{eqnarray}
D{d^2f_i\over dy^2}=S_f^i \; , \label{eq:ion}
\end{eqnarray}
where $S_f^i$ is the net sink term (including collisional
ionization, radiative and dielectronic recombination, charge
transfer, etc., and the reverse processes which populate the current
species) of the species $i$. The turbulent diffusivity, $D$, is of
order unity and assumed to be position-independent. At the inner and
outer boundaries of the mixing layer, the ionization fractions are
set by the equilibrium values.

To complete the description of the mixing layer, we require lateral
pressure equilibrium (which determines the density of the flow along
$y$), and calculate the turbulent viscosity with a simple, mixing
length parametrization of the form:
\begin{eqnarray}
\mu=\alpha\,{\rhol}\,\cl\,\lo \; ,\label{eq:mu}
\end{eqnarray}
where \rhol\ and \cl\ are the mass density (\gcc) and sound speed
(respectively) averaged over the cross-section \lo\ of the mixing
layer (Fig.~\ref{fig:dia}). From the work of Cant\'o \& Raga (1991),
the value of the proportionality constant $\alpha$ is $0.00247$, as
described in Appendix\,A of \pap. It is the required value for a
supersonic mixing layer model to match the opening angle of $\approx
11^{\circ}$ of subsonic, high Reynolds number laboratory mixing
layers in the limit in which the jet Mach number tends to one.

Considering that the turbulent conduction and diffusion Prandtl
numbers are of order one, we can compute the conduction coefficient
as $\kappa\approx \mu c_p$ (where $c_p$ is the heat capacity per
unit mass averaged across the mixing layer cross-section) and the
diffusion coefficient as $D\approx \mu/\rhol$. In this way, we
obtain a closed set of second order differential equations
(\ref{eq:flo} and \ref{eq:ion}), which can be integrated with a
simple, successive overrelaxation numerical scheme.

\subsection{TML calculations with the multipurpose code \map} \label{sec:map}

We use the code \map\ (Ferruit et\,al. 1997) to compute the
radiative energy loss term $L$ and the  photoheating term $G$
(Eqn.\,\ref{eq:flo}) at each position across the TML. At both the
inner and outer boundaries of the mixing layer, we assume
equilibrium ionization of the different species, while across the
layer, our simple overrelaxation scheme is used to determine the
ionization fractions (Eqn.\,\ref{eq:ion}). For the ion diffusion of
each species $f_i$, the spatial differential equations are converted
to temporal equations, with the use of pseudo-time steps $\Delta t =
\Delta{y}^2 ~\lo/(\alpha \,{\bar n} \,\cl)$, where ${\bar n}$ is the
average H\,number density. This allows us to use the temporal
algorithm previously described in Binette \& Robinson (1987) for
determining the spatial diffusion of the ionic species.
%The source terms for the atomic/ionic rate equations make use of .

The radiative transfer is determined by integrating (from the hot
layer $y=\lo$ down to $y=0$) the intensity of the UV diffuse field
produced by the layer, assuming the {\it outward only}
approximation. Any UV radiation impinging the layer from the outer
boundary is simply added to the diffuse field at the onset of the
integration. The intensity of the external ionizing field is defined
by the ionization parameter as follows:
\begin{eqnarray}
\Uo= \phiw/c\,\no \; , \label{eq:uo}
\end{eqnarray}
where $c$ is the speed of light, \phiw\ the flux of ionizing photons
impinging on the TML, and \no, the H\,density at $y=0$. The flux
\phiw\ is the sum of two components: \phis\ resulting from external
UV sources such as hot stars and \phix\ due to the X-ray diffuse
field generated by the hot wind (of column \Nx\ and average
temperature \Tx). In circumstances where the hot wind occupies a
large volume, the soft X-ray luminosity of the hot wind can become
significant despite its density, which is much lower than the
optical line emitting gas. In this paper, we neglected the term
\phix\ because the hard radiation appears to play a negligible role
in comparison with the UV from the hot stars. In fact, using the
diagnostic line ratio diagrams of Veilleux \& Osterbrock (1987), we
find that \ngc\ occupies the same locus as that of \hii-region like
galaxies (DUCBR, Gonzalez-Delgado et\,al. 1994).
%Because the observed emission spectrum is unambiguously produced by young stars,
%therefore neglected the term \phix\ by setting \Nx\ to a negligible value.

The impact of photoionization can be inferred from the behavior of
\gam\ across the layer, where \gam\ is defined as follows:
\begin{eqnarray}
\gam={{L-G}\over{L+G}} \; .\label{eq:gam}
\end{eqnarray}
The quantity \gam\ is zero when the temperature corresponds to the
equilibrium value and it is unity when cooling dominates. In our
calculations, we iteratively determine \To\ at $y=0$ until the
condition $\gam=0$ is satisfied.
%The absorption processes that are taken into account in the transfer
%equations across the TML include all the relevant photoelectric
%cross sections of the ions present.

For the calculations presented in this work, the mixing layers occur
at the outer surface of gas condensations immersed in the radiation
field of hot stars permeating supercluster\,A of \ngc. The spectral
energy distribution (\sed) that we adopted for \phis\ was calculated
using the code LavalSB (Dionne \& Robert 2006). It corresponds to a
newly formed star cluster 1\,Myr old, in agreement with the upper
age limit derived by Drissen et\,al. (2000). The stellar masses are
represented by a Salpeter distribution  with an upper mass cut-off
of 100\,\msol.  The abundances of the atomic elements of the gas are
set at 20\% solar, in line with the conclusions reached by
Luridiana\footnote{The abundances inferred by these authors are 25\%
solar for the stellar atmospheres and 20\% for the nebular gas.}
et\,al. (1999).

After computing the emission line spectrum of a given TML, \map\
offers the option of computing separately the emission lines
generated by the inner photoionized layer (i.e. $y<0$ in
Fig.\,\ref{fig:dia}) where equilibrium ionization prevails. A simple
isobaric photoionization model is calculated in this case using the
radiation field that has \emph{not} been absorbed by the mixing
layer. The total line spectrum is then given by taking the sum of
the line intensities from the TML model (the broad profile
component) and from the static photoionization model (the narrow
profile component).

Since the mixing layer is isobaric,its density profile as a function
of thickness is uniquely determined by the pressure \Po. Given the
value of \no\ at the boundary $y=0$, the pressure is derived from
\To, which is the equilibrium temperature in the photoionized case.
Within our selected $1.2\arcsec\times1.2\arcsec$ aperture
(Sect.\,\ref{sec:sim}), the observed \sii\ (6716/6731) ratio is
1.36, which translates into a density\footnote{This density only
applies to the photoionized layer of the condensations. The core of
the condensations are expected to be cold and therefore much
denser.} of 89\,\cc, assuming a temperature of 12\,000\,$^{\circ}$K.
Within a factor of two, this density is also representative of the
layers that emit \nii\ or \oii. Thus all models presented in this
work were calculated assuming \no=100\,\cmc. It is much lower than
the critical density of most atomic transitions typical of TMLs,
which means that the gas density {\it per\,se} is not a significant
parameter in these calculations. In this case, equal \emph{external}
parameters can be considered equivalent whenever the product of the
H\,density \no\ and thickness \lo\ is the same. Therefore, it is
sufficient to specify the quantity $\No=\no\lo$ to uniquely define a
model, when \Uo\ is kept constant.
%The actual width is then directly inferred from any
%particular density one chooses to select.

To  summarize, in order to compute solutions to the mixing layer, we
must specify the values of the following parameters: the ionization
parameter, the mixing layer's nominal column\footnote{The quantity
$\No\ (=\no\lo$) is a convenient model descriptor. However, it is a
bad estimator of the true integrated H\,column, which is a lot
smaller since the density is not constant but decreases as the
temperature rises with thickness $y$ (Fig.\,\ref{fig:gam}).} \No\
and finally the temperature and velocity of the hot wind: \Tw\ and
\vw.
%\vw\ can be expressed in terms of the wind Mach number,
%$\Mw=\vw/\co$, where \co\ is the adiabatic sound speed in the \emph{static} layer.

\subsection{A high-velocity hot stellar wind}\label{sec:wind}

\begin{figure}
\resizebox{\hsize}{!}{\includegraphics{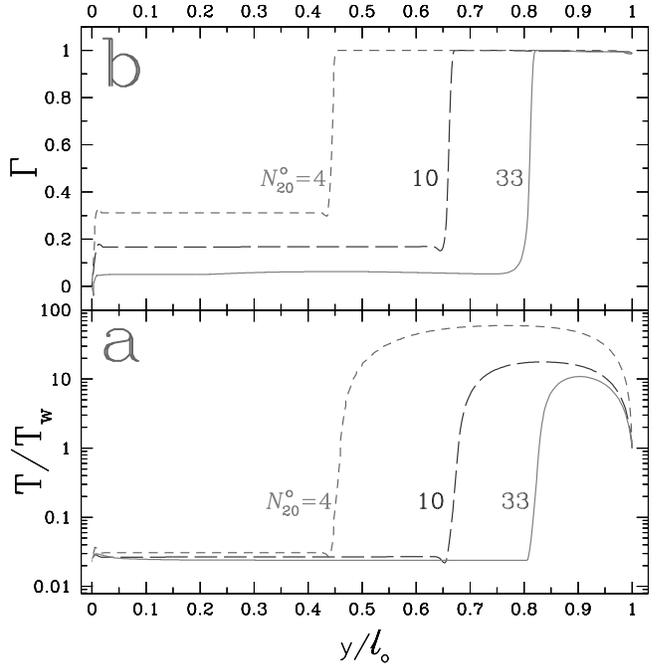}}
\caption{Behavior in panel\,\emph{a} of the temperature normalized
to the value in the wind and in panel\,\emph{b} of \gam\
(Eqn.\,\ref{eq:gam}), both as a function of the \emph{normalized}
thickness $y/\lo$. The three models shown vary only by their mixing
layer nominal column: $\Ntw=4$, 10 and 33.
%and 10\% of the impinging ionizing \sed\ is absorbed within the TML.
%A static ionization-bounded photoionized slab (without TMLs) would result
%in a total column of $\Ne\sim 1.2\times 10^{18}$\,\cms.
} \label{fig:gam}
\end{figure}

The full-width at zero intensity (\fwzi) of the \ha\ and \oiii\
profiles reaches the remarkable value of 7000\,\kms\ in \ngc. This
means that for a geometry in which the emitting gas moves radially
in 3D, the velocity must extend at least up to 3500\,\kms. In
practice, faster winds are required in our models, because the
layers that have velocities approaching that of the wind do not
produce any optical lines. The main reason is that these layers are
very hot and their densities so low that their line emissivities
become negligible. Another reason is that these gas layers are
overionized. Towards the static layers, the temperature is close to
being isothermal because photoheating equals radiative cooling. The
transition between the isothermal layers and the wind dominated
layers is quite abrupt. Therefore the thermal structure of TMLs
consists of two zones: the isothermal warm zone and the hot thermal
bump. Examples of such a thermal structure is illustrated in
Fig.\,\ref{fig:gam}\emph{a} as a function of the \emph{normalized}
thickness $y/\lo$. The three models shown are described in detail in
Sect.\,\ref{sec:stru} below and differ only by their value of \No\
(hereafter \Ntw, in units of $10^{20}$\,\cms). The normalized
thickness maps directly into the velocity domain, since the Couette
flow solution implies a linear velocity increase between the static
layer and the wind flow (Eqn.\,\ref{eq:vy}). If we define \ep\ as
the fraction of the layers' thickness where \Tw\ is isothermal,
which turns out to be where the optical lines are produced, it
follows that a wind velocity of $3500\, \ep^{-1}$\,\kms\ is required
to cover the velocity span observed in \ngc. We could not get
credible calculations that had $\ep \ge 0.9$. The model with
$\No=33$ in Fig.\,\ref{fig:gam}\emph{a} has $\ep=0.81$. Hence, in
order to reproduce the observed broad wings, it is necessary to
adopt a wind velocity as high as $\vw \simeq 4300$\,\kms. This value
is somewhat higher than the 3500\,\kms\ value calculated by
Sternberg et\,al. (2003) for O3 stars.

Rather than a single stellar wind, we propose that we have a wind
from a dense cluster of massive stars. As shown by Cant\'o et\,al.
(2000), the resulting cluster wind has a terminal velocity equal to
the velocity of the winds from the cluster stars. Also, the
interaction of the different winds results in a very high initial
temperature for the cluster wind ($\sim 10^8\,^{\circ}$K for the
wind velocity we are proposing), but this temperature rapidly drops
beyond the outer radius of the cluster, as the cluster wind
approaches its terminal velocity.

Another alternative to a stellar origin for the wind is that it has
been generated by supernovae. There is, however,  no evidence of
past supernovae in knot\,A although we cannot entirely rule it out.
In knot\,B, there is evidence of a cavity surrounding the central
stellar cluster, which might have been caused by supernovae
explosions. The explanation provided by Drissen et\,al. (2000, 2001)
is that supercluster\,A is extremely young ($\la 1$\,Myr) while
supercluster\,B is older ($\sim 3$--4\,Myr).

We found that the temperature \Tw\ of the hot wind is not a critical
parameter of the TML models. Its value can be varied and still
result in a sequence of equivalent models (i.e. with similar line
ratios), provided \vw\ remains the same. However, to ensure that the
X-ray flux from the hot wind would not be excessive, we have set its
temperature to $\Tw=5 \times 10^{5}\, ^{\circ}$K in all our
calculations. The non-detection of \ngc\ by the satellite
\emph{XMM-Newton} in the 0.1--10\,keV X-ray band  has allowed the
determination of an upper limit of $4\times 10^{-13}$\,\flu\ (Y.
Krongold \& E. Jimenez-Bailon, private communication). Using the
code MEKAL (Mewe et\,al. 1986; Liedahl et\,al. 1995) and assuming a
cylinder of hot gas of radius 10\,pc, length 100\,pc and density
$\nw= 100\,\cmc \times 12\,000\,^{\circ}{\rm K}/\Tw=2.4\,\cmc$, one
of us (Y. K.) recently determined that the wind would emit a flux
more than an order of magnitude lower than the above upper limit.
Thus, even much higher wind temperatures ($\sim 10^8\,^{\circ}$K)
would neither produce a flux that exceeded this upper limit.

%of a cluster wind flow that results from the multiple interaction of
%the stellar winds produced by a dense cluster of massive stars.

%\Mw\ is scaled proportionally to $\Tw^{-0.5}$ (this is equivalent to
%conserving the same values of the wind \vw).

%$\cso=(\frac{5}{3}k\To/\mu m_H)^{0.5}$
%with the following values: (He, C, N, O, Ne, Mg, Si, S, Ca,
%Ar, Fe) $= [10^5, 363, 112, 851, 123, 38, 36, 16, 2.3, 3.6, 47]
%\times 10^{-7}$ relative to H (by number).

\subsection{Internal structure of three test models}\label{sec:stru}

After defining the stellar \sed, density and wind characteristics,
we can now proceed with the calculations. In order that the line
ratios of the photoionized layers could match the nebular values, we
selected a high ionization parameter $\Uo=0.02$. The total
\oiii/\hb\ ratio, including the static gas emission, is $\approx
8.1$, which is comparable to the observed ratio of 8.9. Our main aim
is to reproduce the broad lines profiles and therefore no further
analysis of the line ratios of the cores of the profiles has been
attempted. In Fig.\,\ref{fig:gam}, apart from \Tw, we also show the
behavior of the measured imbalance between cooling and heating,
\gam\ (see Eqn.\,\ref{eq:gam}), as a function of normalized
thickness for each of the three models whose mixing layer's column
differ and take on values of $\Ntw=4$, 10 and 33, in units of
$10^{20}$\,\cms. For each model, the ionization parameter is
$\Uo=0.02$ and the wind temperature and velocity are $\Tw=5\times
10^{5}\,^{\circ}$K and $\vw=4300$\,\kms, respectively.
%for an 80\% absorbed external \sed\ from a 38\,000\,K stellar atmosphere.
The equilibrium temperatures at the onset of the layer are
$\To=11\,460$, 11\,330 and 12\,370\,$^{\circ}$K for the models with
$\Ntw=4$, 10 and 33, respectively. Because \ep\ increases with \Ntw,
these TML models result in a broad profile whose width increases
monotonically with \Ntw. Alternatively, a sequence of increasing
profile's width can be obtained by increasing \vw\ while keeping
\Ntw\ constant.

It is well known that the cooling rate $L$ of a hot plasma decreases
with temperature when its value lies above $\approx 2\times 10^{5}\,
^{\circ}$K (e.g. Fig.\,8 in Ferruit et\,al. 1997). This behavior of
the cooling is responsible for the temperature bump towards the
interface with the wind, which is apparent in Fig.\,\ref{fig:gam}.
In effect, near the wind interface, the heating from turbulent
dissipation overwhelms radiative cooling and results in \gam\ near
unity. It also results in an overionized plasma of very low density
(hence low emissivity), which does not generate significant line
emission, at least in the optical domain. The impact of heating by
turbulent dissipation increases considerably when we consider
thinner mixing layers, as can be seen in Fig.\,\ref{fig:gam}.
Towards the colder gas layers, radiative cooling becomes
sufficiently strong to enable a balance between radiative cooling
and heating by photoionization and turbulent dissipation. Although
heating by photoionization dominates in the near-isothermal region,
turbulent dissipation is nevertheless present as indicated by \gam,
which reaches 0.3 in the case of the $\Ntw=4$ model.

\subsection{Projection in 3D and line profile calculations}\label{sec:pro}

\begin{figure}
\resizebox{\hsize}{!}{\includegraphics{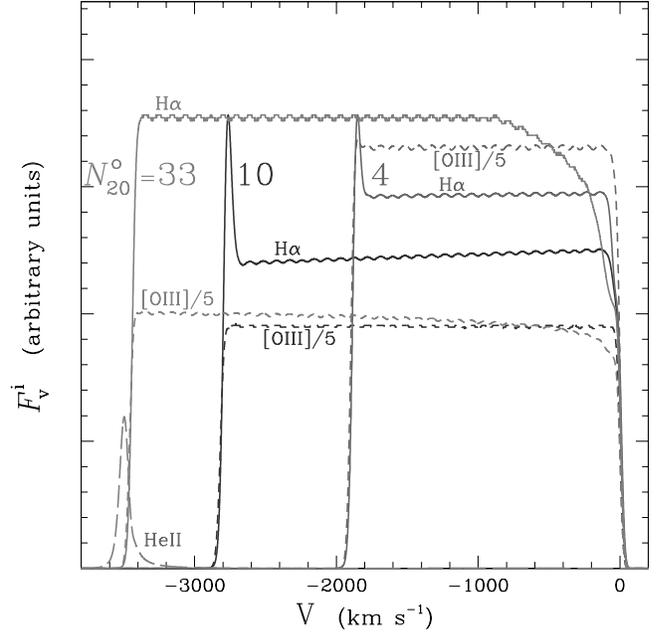}}
\caption{Emission line flux from \ha\ (solid line) and \oiiiw\
(dashed line) as a function of radial velocity is plotted for each
of the three models of Fig.\,\ref{fig:gam}. Note that the \oiii\
fluxes have been scaled down by a factor 5. The \heiiw\ emission
flux is shown in the case of the \Ntw=33 model only.  }
\label{fig:emi}
\end{figure}

In order to derive the integrated line profiles from the 3D
distributions of gas condensations, we first compute the profile
from a single condensation at the surface of which the fast wind
generates a TML. We neglect the details of the contours of such
condensation and consider that the TML takes place on the sides of
the condensation, as if it was a cylinder whose axis is oriented
radially with respect to the wind source. In Fig.\,\ref{fig:emi} we
show the resulting emission flux from the same three models of
Sect.\,\ref{sec:stru}, as a function of radial velocity. The
vertical scale is arbitrary between models, but remains the same for
lines of the same model. These 1D profiles correspond to a single
condensation that lies exactly along our line-of-sight to the wind
source. The profiles have been initially smoothed in \map\ by a
narrow gaussian of 25\,\kms\ dispersion (see Sect.\,\ref{sec:app}),
which accounts for the virial broadening due to the supersonic
motions of the condensations. To a first order and for most optical
lines, the profiles resemble a top-hat function that extends up to a
limit, \vli. The velocity limits up to which the \ha\ or \oiii\ flux
extends are $\vli= -1900$, $-$2800 and $-$3440\,\kms\ for the models
with $\Ntw=4$, 10 and 33, respectively.  This top-hat shape is the
result of the Couette flow solution, which corresponds to a velocity
that increases linearly with $y$ (Eqn.\,\ref{eq:vy}). The \heii\
line is an exception because it is mainly produced at the onset of
the thermal bump. \vli\ is lower than \vw\ since the optical line
emissivities become negligible inside the overionized thermal bump
region, as discussed above. Essentially, all of the flux originates
from the near-isothermal region (depicted in Fig.\,\ref{fig:gam}).
The thinner the TMLs gets, the smaller becomes the region where
emission takes place, which results in proportionally narrower line
profiles. Furthermore, the model with \No=33 is ionization-bounded,
while the other two with \No=10 and 4 are matter-bounded,  which
result in significantly higher \oiii/\hb\ line ratios, of 10.8 and
15.7, respectively.

The emission profiles of Fig.\,\ref{fig:emi} apply to the simple
one-dimension case. We will now consider the more realistic case of
a radial wind in 3D and compute the profile that would be seen by a
distant observer. To achieve this, we assume some simplifications
about the problem, namely, that all the TMLs are immersed in a
radial constant velocity wind, the wind is isotropic with respect to
the wind source and the covering factor $C_f$ of the mixing layers
is also isotropic. To simplify things further, we will assume that
the TMLs are located at the same radius $R$ from the central source
of the wind. For any emission line $i$ at rest-wavelength
$\lambda_o^i$, these considerations lead to the following integral
for the flux:
\begin{eqnarray}
\begin{array}{rlllll}
F_v^i &= 2\pi R^2 C_{f}  \left\{\mathlarger{\int_{\thei}^{\thef}}
F^i\!(v^{\prime})
\sin\theta d\theta \right. \\
+& \left.
\mathlarger{\int_{\frac{\pi}{2}+\thei}^{\frac{\pi}{2}+\thef}}
F^i\!(-v^{\prime}) \, \exp\!\left(-\tau_{\lambda_o^i} \cos \theta\right) \sin \theta d\theta \right\} , \label{eq:pro}\\
F^i\!(v^\prime) &= \sec\theta \,F^i\!(v\cos\theta) \; ,\\
v^\prime &= v \cos\theta \; ,\\
\tau_{\lambda_o^i} &= \sigma_d^{i} \,n_{\!d} \,2 R \; ,\\
v &= c(\lambda-\lambda_o^i)/\lambda_o^i \; ,
\end{array}
\end{eqnarray}
where $c$ is the speed of light, $\theta$ the angle with respect to
the line-of-sight to the observer, $R$ the radius where the TMLs are
located with respect to the wind source, and $F^i\!(v)$ in the
integrand is the emission flux for line $i$ as computed by \map\
(i.e. for $\theta=0$). To include the possibility that dust might
fill the volume of the sphere of radius $R$ and thereby absorb line
emission from the (red-shifted) TMLs on the far-side, we introduce
in Eqn.\,\ref{eq:pro} a dust extinction cross-section $\sigma_d^{i}$
evaluated at $\lambda_o^i$. The parameter $n_{\!d}$ is the dust
grain density filling the sphere and $\tau_{\lambda_o^i}$ is the
resulting line extinction due to dust. The effect of internal dust
would be to skew the profile towards the blue. Since we found no
evidence of profile asymmetry, we have set $n_{\!d}=0$. There could
be additional `intervening' dust that covers the whole nebula,
however. This possibility is not relevant to our profile study since
the differential reddening across the profile width turns out to be
negligible, assuming a \chb\ of 0.2 for knot\,A (Gonzalez-Delgado
et\,al. 1994). As for the two limits of integration, whenever the
TML system (of projected area $\pi{R}^2$ on the sky) is fully
contained within the aperture of the spectrograph, these reduce to
$\thei=0$ and $\thef=\pi/2$, which is what is assumed hereafter. We
have explored relaxing this assumption. With $\thef\la\pi/2$, the
profiles become flat-topped while for $\thei\ga{0}$, the profiles
does not change much in shape but becomes progressively narrower. In
a more realistic description, the TMLs would cover a range in radii
and $C_f$ would be a function of radius. The current description,
however, suffices to capture the basic implication of a 3D geometry
on the line profiles.

\section{Comparison of TML profiles with the
observations}\label{sec:sup}

\subsection{Characteristics of the broad profile wings}\label{sec:sim}

We analyzed a subset of the available data as follows. To derive the
highest S/N possible, we considered a square region of
$1.2\arcsec\times1.2\arcsec$ ($20\times20$\,pc) centered on knot\,A
and extracted a red and a blue spatially-collapsed spectra within
this area.  We modeled the underlying continuum and subtracted it
from both the \ha\ and \oiiiw\ lines. In order to facilitate the
comparison of the models with the faint broad wings observed in
\oiii\ and \ha, we converted the profiles into velocity space
relative to the centroid of each line. Direct profile comparison is
achieved by simply superposing different lines.

%We found that if we scaled \oiii\ vertically so that its narrow core
%had the same flux as that of \ha, the broad wings of \ha\ lied higher by about 50\%.

\begin{figure}
\resizebox{\hsize}{!}{\includegraphics{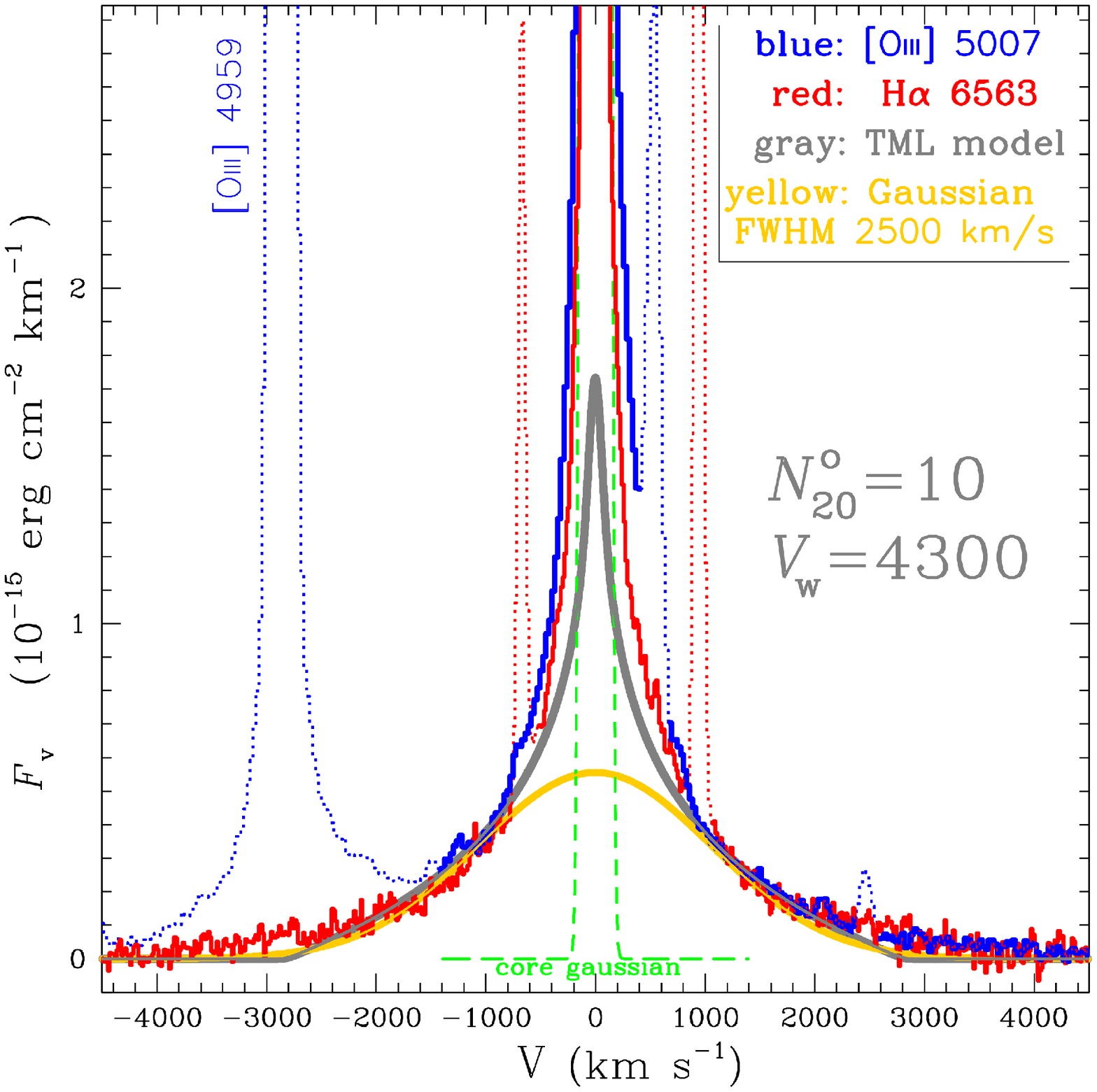}}
\caption{Superposition of the emission line profiles of \ha\ (red
line) and of \oiiiw\ (blue line) as a function of Doppler velocity
with respect to the centroid of both lines. The scale applies to
\ha\ and the units are \fvun. The \oiii\ profile has been scaled up
%by a factor by 1.5
so that its broad wings coincided as much as possible with those of
\ha. Emission lines involving other energy levels have been
partially blanketed with dotted line contours. Using a thick gray
line, we overlaid the 3D profile from a TML model with \Ntw=10 and
\vw=4300\,\kms. The yellow line shows how a broad Gaussian of
2500\,\kms\ \fwhm\ would compare. The green colored dashed-line is a
\fwhm=110\,\kms\ Gaussian fit of the narrow core component.}
\label{fig:pro}
\end{figure}

At the time of writing, various issues concerning the absolute
calibration of the blue spectrum could not be satisfactorily
resolved. In what follows, we will not rely on the absolute flux
scale of the blue spectrum, but focus instead on profile shape
comparisons or on relative ratios of the broad component with
respect to the central core component. After rescaling the blue
spectrum until the broad \oiii\ superimposes the broad \ha\ profile,
one finds that the `shape' of the broad component is the same in
both lines. This is apparent in Fig.\,\ref{fig:pro}, where the \ha\
profile as a function of Doppler velocity is shown in red and
\oiiiw\ in blue. To express fluxes, the quantity $\fv=10^{15}F_{v}$
is used hereafter. The \ha\ line peaks at a value of \fv=853 (but
not \oiii\, which has been rescaled in this figure). No reddening
correction has been applied. We now discuss plausible
interpretations of the scaling factors that are obtained from
superimposing line profiles.

Detailed analysis showed that if we scaled the \oiii\ line profile
so that the peak of its narrow core equalled that of \ha, the broad
wings of \ha\ were brighter than \oiii\ by about 50\%. If instead of
\oiii, we overlaid the continuum subtracted \hb\ line and rescaled
it so that the peak of its narrow core equalled that of \ha, we
similarly found that the broad \ha\ lies $\approx 50$\% higher than
the broad \hb. We verified that saturation is not taking place. Flux
spilling over nearby pixels appears also to be ruled out. A possible
interpretation is that the gas responsible for the faint broad wings
is further absorbed by dust than the gas responsible for the bright
narrow core\footnote{The amount of dust encountered by
Gonzalez-Delgado et\,al. (1994) for the narrow line emitting gas in
knot\,A is as little as \chb=0.2 (or $\tau_{\rm V}= 0.22$).
Uncertainties in our absolute calibration of the blue spectrum does
not allow us to verify this value. To have more dust covering the
broad line emitting gas, however, is somewhat counterintuitive and
this particular interpretation should be considered tentative.}.
%so we could not find an alternative explanation.
Apart from our basic conclusion that the broad \oiii\ and \ha\
profiles are of similar shapes, which suggests that the broadening
mechanism acts uniformly on both lines, we find that the data is
consistent with the broad \oiii/\hb\ flux ratio being the same as
that of the narrow core \oiii/\hb. This follows from the two
separate comparisons above that indicated that the broad \hb\ and
\oiii\ are depressed by the same factor with respect to the broad
\ha. It is noteworthy that these two conclusions are not affected by
our calibration uncertainties.

Hereafter we will concentrate on the faint broad profiles shown in
Fig.\,\ref{fig:pro}. As discussed above, the \oiii, \ha\ and \hb\
broad profiles are characterized by a similar functional dependence
on velocity. This similarity in shape  and the fact that the
integrated \oiii/\hb\ flux ratio for the wings is the same as for
the narrow core imply stringent constraint for any proposed model.
It suggests that the physical conditions pertaining to the gas that
produces the faint broad wings are very likely similar to those of
the more quiescent nebular gas.

The \fwhm\ of the \emph{core} of the profiles was derived by fitting
a single Gaussian. We obtained 78 and 115\,\kms\ for \ha\ and \oiii,
respectively. The \oiii\ core profile is unresolved and appears
broader due to the lower resolution of the blue grating. In order to
distinguish at which velocity the broad line flux exceeds the
intense flux from the narrow core, we overlay in Fig.\,\ref{fig:pro}
\,a Gaussian fit (green dashed line) to the narrow core (with
\fwhm=110\,\kms). We infer that the observed profiles becomes
noticeably wider than the narrow Gaussian below $\fv\approx2$, which
lies at only 0.23\% of the \emph{peak} flux value. Note that the
width of the narrow Gaussian is 325\,\kms\ at \fv=2.
%\com{check broad component affected or not by same reddening as narrow component}{lb}
%To guide the eye, a narrow Gaussian of \fwhm=110\,\kms\ representing the core
%component is shown (green colored dashed-line).

\subsection{Comparison with 3D TML line profiles}\label{sec:com}

Since our main aim is to reproduce the weak broad wings underneath
each line, we now focus on the profiles generated by the TML,
leaving out the narrow line component produced by the \emph{static}
layer at the surface of the condensations.
%each of which can be moving supersonically relative to the other condensations.
Since the emission flux from the broad component is proportional to
the surface area of the TML gas exposed to the ionizing radiation,
this area has to be quite small ($\approx 3$\%) relative to that of
the nebular gas (the `static' or narrow line component).
%since our main aim is to reproduce the weak broad wings underneath each line.

\begin{figure}
\resizebox{\hsize}{!}{\includegraphics{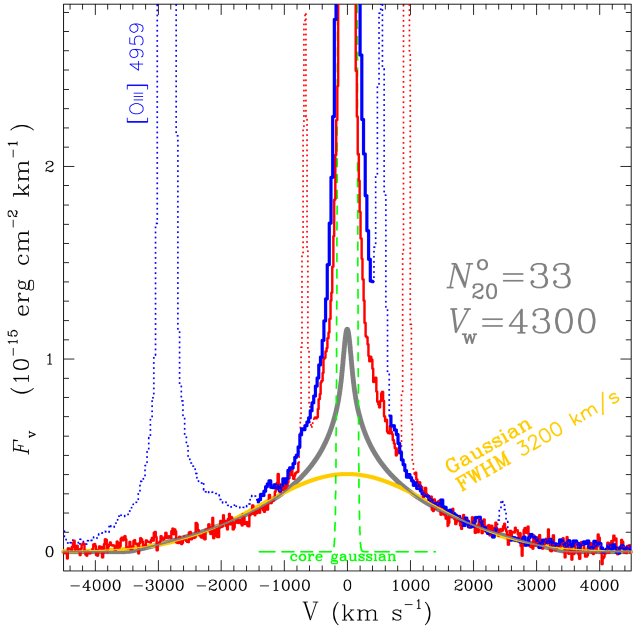}} \caption{Same
notation as in Fig.\,\ref{fig:pro} except for the TML model, which
corresponds to \Ntw=33. For comparison, a Gaussian of 3200\,\kms\
\fwhm\ is overlayed. } \label{fig:proo}
\end{figure}

\begin{figure}
\resizebox{\hsize}{!}{\includegraphics{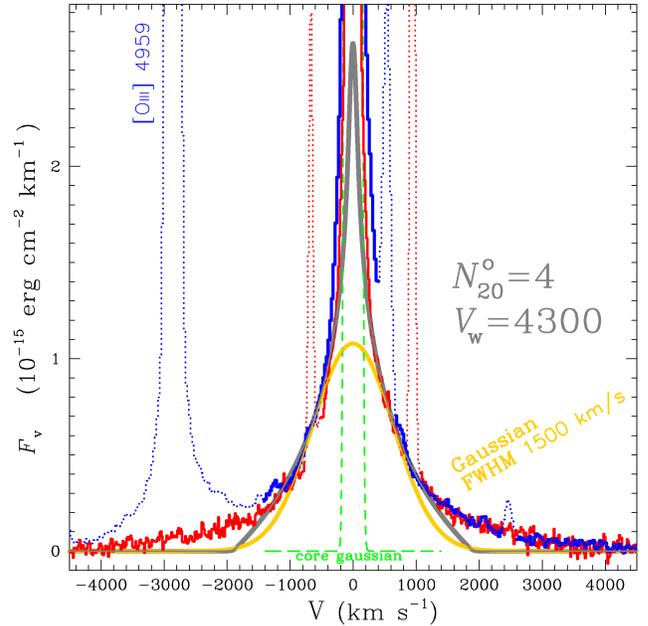}} \caption{Same
notation as in Fig.\,\ref{fig:pro} except for the TML model, which
corresponds to \Ntw=4. For comparison, a Gaussian of 1500\,\kms\
\fwhm\ is overlayed. } \label{fig:prooo}
\end{figure}

We applied the profile integral described in Eqn.\,\ref{eq:pro},
which considers the effects of velocity projection in 3D, to each of
the three TML models shown in Fig.\,\ref{fig:emi}. Because the
narrow line cores in \ngc\ have a width that exceed the value
assumed by \map\ (25\,\kms, see Sec.\,\ref{sec:app}), we further
convolved\footnote{The target width is 115\,\kms, which implies a
convolution by a Gaussian of \fwhm\ of
$\sqrt{115^2-25^2}=107$\,\kms.} the calculated profiles using a
narrow Gaussian of 107\,\kms\ \fwhm. The \fwhm\ describing the
velocity dispersion of the condensations (Fig.\,\ref{fig:cart}) also
applies to the TMLs.

\begin{figure}
\resizebox{\hsize}{!}{\includegraphics{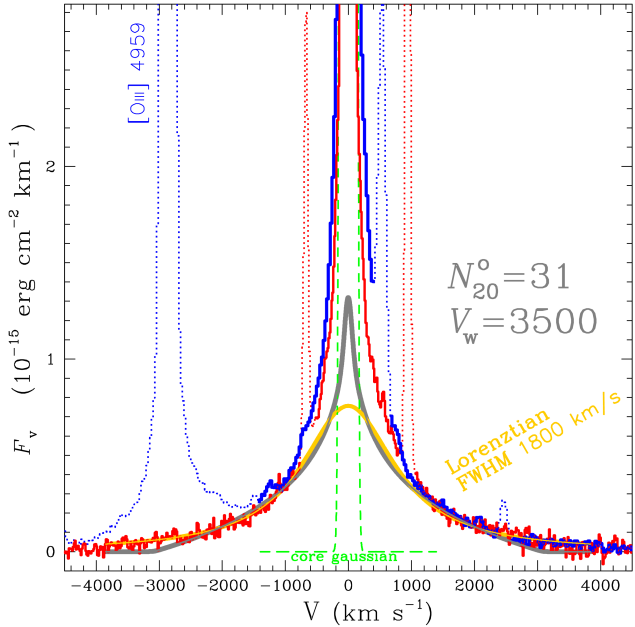}} \caption{Same
notation as in Fig.\,\ref{fig:pro} except for the TML model, which
corresponds to a slower wind of \vw=3500\,\kms\ with \Ntw=31. For
comparison, a Lorentzian profile of 1800\,\kms\ \fwhm\ is overlayed.
} \label{fig:proooo}
\end{figure}

The TML model of the \oiii\ line assuming $\Ntw=10$ is superimposed
to the data in Fig.\,\ref{fig:pro} (gray line). It is unnecessary to
superimpose the \ha\ model separately, since it is visually
undistinguishable from the \oiii\ model, even if the difference in
spectral resolution were considered in the modeling. The profile fit
that is achieved of the broad wings, although clearly imperfect, is
certainly encouraging, given the approximations made to the transfer
and to the geometry of the TMLs. Because Gaussians are a natural
profile for describing emission lines over a wide variety of
physical situations, we also compared our data with simple Gaussian
profiles. In Fig.\,\ref{fig:pro}, the yellow line represents a
Gaussian with an \fwhm\ of 2500\,\kms. We find that our TML model
provides a better description of the data. (Gaussians with different
\fwhm\ are drawn in the subsequent Figures \ref{fig:proo} and
\ref{fig:prooo}). There is definitely more flux present at both low
and high velocities in the \emph{observed} profiles than provided by
any single broad `Gaussian'. The same appears to be the case for the
TML model with \No=10, but to a much lesser extent. For instance,
the TML profile fills the widening of the profile beyond the narrow
Gaussian core (dashed green line) much better compared to the round
top of a wide Gaussian. A Lorentzian profile would provide a better
description of the data than a Gaussian, as shown in
Fig.\,\ref{fig:proooo}. However, given the physical characteristics
of the gas, there is no physical justification for a Lorentzian
profile.

In Figs.\,\ref{fig:proo} and \ref{fig:prooo}, we show the other two
TML models of equal \vw, but with different \Ntw\ of 33 and 4,
respectively. The model that fits the wings better
(Fig.\,\ref{fig:proo}) does worse with the core region, and
conversely with the other model. For comparison, Gaussians that
approximately match the same region as the corresponding TML model
are overlaid.

To obtain a much improved fit to the profile wings, the
juxtaposition of two or more TML profiles of different widths would
therefore be necessary. Because the \oiii/\hb\ ratio varies markedly
along the \No\ sequence (Sect.\,\ref{sec:pro}) and since the
observed broad \oiii/\hb\ line ratio is the same as the one found
for the narrow core, we would argue against combining models of
different thickness \No.  An alternative way to combine TML profiles
would be to consider a radial gradient in wind velocity \vw.  Such a
sequence in which \vw\ varied and \No\ remained at the value set by
the ionization-bounded case would present a clear advantage, since
the resulting \oiii/\hb\ ratio would not change from the value
characterizing the static nebula case.

To illustrate this, we looked for an ionization-bounded model that
would be equivalent to the model with \Ntw=10 (Fig.\,\ref{fig:pro},
which is the model that fits relatively well the bulk of the broad
profile). In this new model, the increase in thickness is
compensated by a reduction in wind velocity. The resulting profile
is shown in Fig.\,\ref{fig:proooo} and corresponds to a slower wind
of \vw=3500\,\kms\ and a larger thickness \Ntw=31. The comparison of
Fig.\,\ref{fig:proooo} with Fig.\,\ref{fig:pro} shows that the two
profiles are comparable. In the case of the slower wind model, the
broad profile is slightly wider because \ep\ is larger (0.88) and it
fares better in the wings than the 4300\,\kms\ model with \Ntw=10.
This new model is preferable, since it results in the same
\oiii/\hb\ ratio as that of the narrow core. Furthermore, the wind
velocity is more conservative and corresponds to the value
calculated by Sternberg et\,al. (2003) for O3 stars.

%in order to cover the breath of the broad profile observed.

\subsection{Other emission lines}\label{sec:high}

The TML models predict that there should not be any broad wings
underneath the lower excitation lines such as \nii\ or \sii. The
reason is that in models with a high ionization parameter (e.g.
$\Uo\ga 0.01$) the degree of ionization throughout the turbulent
layer is high and the ionized gas is dominated by high excitation
species, such as \opp\ or \npp. This means that low excitation
species are relatively insignificant throughout the layer.
%Eventhough the S/N is much lower for the low excitation lines since they are much weaker,
Inspection of the data shows no evidence of broad wings underneath
the low excitation lines at a level comparable to that observed in
\ha, \hb\ or \oiii. We have also explored whether faint wings would
be present underneath the \emph{high} excitation lines, as is
predicted by the model. The limited S/N available for the weak
\ariv\ lines, however, has made such a test inconclusive.

\begin{figure}\resizebox{\hsize}{!}{\includegraphics{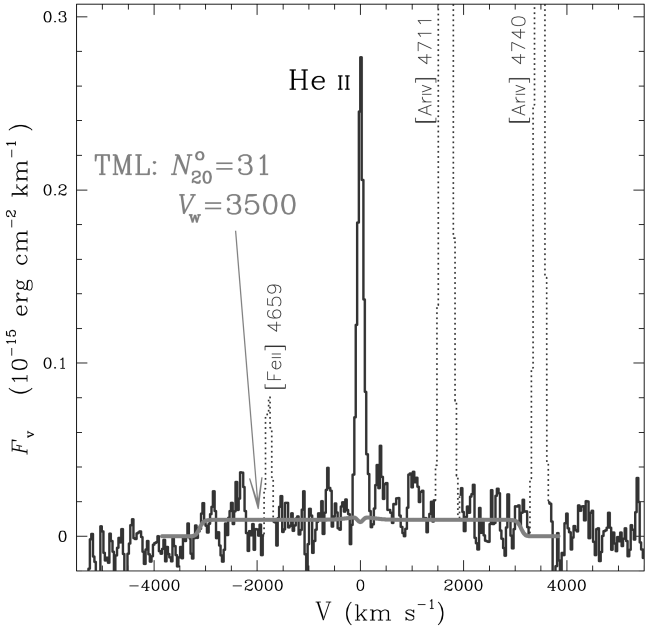}}
\caption{Observed emission line profiles of \heiiw\ (black line) as
a function of Doppler velocity with respect to the centroid of both
lines.  Other line transitions near \heii\ have been blanketed by
dotting the corresponding line.  We overlay a 3D profile from a TML
model (gray line) with \vw=3500\,\kms\ and \Ntw=31, but with its
intensity multiplied by a factor 5.
%This model assumes the same input parameters as in the previous figure (\vw=3500\,\kms).
} \label{fig:heii}
\end{figure}

The shapes of the profiles predicted by the TML models are either
characterized by a broad profile with a peaked center (\oiii, \civ,
\ciii, \neiii, \ha, \hb) or a narrow core without any significant
wings (\sii, \mgii, \nii, \oii). There are a few exceptions,
however, such as the high excitation line of \heiiw, which presents
a top hat shape. This arises as the line is only produced
significantly at the onset of the thermal bump in
Fig.\,\ref{fig:gam} where the temperature shoots up above that of
the isothermal region\footnote{Since the density decreases inversely
to the temperature, the \heiiw\ line emissivity becomes negligible
towards the very high temperature end.} (see \heii\ emission flux in
Fig.\,\ref{fig:emi}). For illustrative purposes we show the profile
predicted for \heii\ in Fig.\,\ref{fig:heii}. It corresponds to an
upper limit equivalent to 5 times the intensity predicted by
%factor was 20 but the blue spectrum was 4 times too low relative to Halpha
the TML model with \vw=3500\,\kms\ and \Ntw=31. It is possible that
\map\ underpredicts the strength of this line, since it considers
only pure recombination and does not include collisional excitation.
It is interesting to note that one cannot completely rule out the
presence of a weak broad component (or flat pedestal) underneath the
narrow \heii\ profile. We do not expect such a component to be due
to a Wolf-Rayet feature, since supercluster\,A in \ngc\ is much too
young ($< 1$\,Myr; Drissen et\,al. 2000) and, furthermore, no
evidence has been found of broad \niii\ or \civ, which are otherwise
expected if Wolf-Rayet stars were present.

\subsection{Line profiles from starburst galaxies}\label{sec:burst}

A different area where TML emission could play a role are broad
profiles, with $\fwhm$ in the range of 150--400\,\kms, observed in
starburst galaxies, such as NGC\,1569 (Westmoquette et\,al. 2007a,
b, 2008) or M82 (Westmoquette et\,al. 2007c). The requirements for
wind velocities in these objects are modest in appearance, which
implies a much lower wind velocity. However, the \fwhm\ observed may
nevertheless correspond to a significantly higher \vw\ if the wind,
rather than being fully isotropic, tended to project along the plane
of the sky of the observer. For instance, in the case of near
edge-on galaxies, the wind can be funneled above and below the
galactic plane, resulting in a narrower profile because $\thei>0$,
in the notation of Sect.\,\ref{sec:pro}. In contrast to \ngc,
supernovae can play an important role in launching starburst winds.
Another significant difference is that it is the lower excitation
lines like \niiw\ that are observed to be broad. This can only be
accounted for in our model if a sufficiently low ionization
parameter is assumed, which could ensure that the intermediate
excitation lines spread over a significant fraction of the TML's
thickness. A softer ionizing continuum due to high metallicities may
also contribute to the prominence of intermediate excitation lines.
In the case of M82, Westmoquette et\,al. (2007c) report a relatively
low \oiii/\hb\ (5007/4861) ratio of $\simeq 0.32$. New TML
calculations adapted to the context of starburst galaxies would be
required to verify whether the inferred ionization parameter is
sufficiently low for the \nii\ lines to become broad as well.

\section{Conclusions}

Four mechanisms have been explored by Roy et\,al. (1992) to account
for the extreme gas velocities observed in \ngc, namely stellar
winds, Thomson scattering by hot electrons, supernova remnants, and
a superbubble blowout. These authors reported significant problems
with each of these mechanisms and considered them to be
unsatisfactory. They concluded that the broad nebular gas is
probably due to ``very high velocity gas whose origin is, at
present, unknown''. In this paper, we explore the possibility that
the broad emission originates from turbulent mixing at the interface
between a hot cluster wind and more quiescent photoionized
condensations. In our model, all gas phases are in pressure
equilibrium and, because of the high temperature of the wind, the
total ionized gas mass contained in the wind is much lower than
postulated by Roy et\,al. (1992), by a factor $\sim \Tw/10^4$. Since
the upper limits on \Tw\ set by the non-detection in the X-rays
allows temperature values as high as $10^8\,^{\circ}$K
(Sect.\,\ref{sec:wind}), the objection to the wind hypothesis
expressed by Roy et\,al. can be lifted.

The basic input parameters affecting the TML models have been given
appropriate values as follows: we inferred a density of $\simeq
100$\,\cc\ for the nebular gas, using the \sii\ doublet. The
metallicity that was assumed is 20\% solar, as deduced by Luridiana
et\,al. (1999). The \sed\ that we adopted was calculated using the
code LavalSB and corresponds to a newly formed star cluster 1\,Myr
old [the upper age limit derived by Drissen et\,al. (2000)].
Finally, the ionization parameter has been adjusted so that the
calculated \oiii/\hb\ ratio became comparable to the observed value.

In our model, the broad profile results from radial acceleration in
3D of photoionized turbulent gas. About 10--20\% of the acceleration
that takes place occurs at temperatures approaching that of the
wind, and this gas does not contribute to line emission. For this
reason, the wind velocity assumed in the TML model must exceed the
value inferred from the extent of the faint wings (i.e.
$\pm3500$\,\kms). To constrain the models, we adopted the spectral
data set of Drissen et\,al. (2009) that was obtained at the Gemini
observatories.  We report in Sect.\,\ref{sec:com} on the results of
models with $\vw=4300$\,\kms, which can successfully fit the faint
wings when the TML is ionization bounded (\Ntw=33;
Fig.\,\ref{fig:proo}). The bulk of the profile is better fitted if
we reduced the TML thickness to $\Ntw \simeq 10$
(Fig.\,\ref{fig:pro}). This model, however, is matter-bounded and
the \oiii/\hb\ ratio from the TML would exceed that observed in the
core of the lines as well as in the broad component. Our preferred
scenario, therefore, is that the cluster wind undergoes radial
acceleration, but the TML remains ionization-bounded at all radii.
Our slower wind model of 3500\,\kms\ and with \Ntw=31 is as
successful in fitting the bulk of the broad profile
(Fig.\,\ref{fig:proooo}). It is therefore our best model overall.
The relative success of TMLs in fitting broad profiles and in
predicting the absence of broad lines for \nii\ and \sii\ are
significant arguments in favor of the models. The requirement of
wind velocities of order 4000\,\kms\ in order to fit the very faint
wings remain a stumbling block, however, unless supercluster\,A,
because of its extreme youth, is able to generate such physical
conditions. Interestingly, our TML model results in a flat-top
profile for \heiiw, although this characteristics could not be
confirmed with the current data. As a follow-up study, new
calculations adapted to the broad component observed in starburst
galaxies (e.g. Westmoquette et\,al. 2007a, b, c) should be carried
out along the lines developed in Sect.\,\ref{sec:burst}.

Although, in this paper, the accelerating mechanism is assumed to
arise from a stellar wind, alternative mechanisms could be
considered that would lead to a near-constant radial acceleration
and therefore generate profiles equivalent to those shown in this
work. One possibility for instance might be radial acceleration due
to radiation pressure acting on dusty photoionized gas plumes. \ngc\
is clearly a fascinating object that remains a challenge in our
understanding of extremely young nebulae.

%To summarize, TMLs result in profiles that matches the broad wings observed
%in \ngc\ lines reasonably well, much better than a simple Gaussian profile.

\acknowledgements
%We thank Yair Krongold (IA-UNAM) and Elena
%Jimenez-Bailon (IA-UNAM) for working out an X-ray upper flux limit
%from \emph{XMM-Newton} archives.
L.\,B. owes the inspiration of considering TMLs in the context of
star forming regions to M.\,S. Westmoquette. We thank the unknown
referee for his contribution to the clarity of the ideas presented
in this paper. This work was supported by the CONACyT grant J-50296.
L.\,D. acknowledges financial support from the Canada Research Chair
program. CR and LD acknowledges financial support from Canada's
Natural Science and Engineering Research Council (NSERC) and from
Qu\'ebec's ``Fonds qu\'eb\'ecois de la recherche sur la nature et
les technologies'' (FQRNT).
%We are thankful to xxxxx for constructive discussions about the manuscript.
Diethild Starkmeth helped us with proofreading.

%\begin{appendix}
%\section{The derivation of constant $\alpha$}\label{app:a}
%ff gggg blabla
%\end{appendix}


\begin{thebibliography}{}

\bibitem[Begelman \& Fabian 1990]{beg90} Begelman, M. C., \& Fabian,
A. C. 1990, \mnras, 244, 26

%\bibitem[Binette, Dopita \& Tuohy 1985]{bin85} Binette, L.,
%Dopita, M.A.,  Tuohy, I.R., 1985, ApJ 297, 476

%\bibitem[Binette et al.(1993)]{1993ApJ...414..535B} Binette, L., Wang, J.,
%Villar-Martin, M., Martin, P.~G., \& Magris C., G.\ 1993, \apj, 414, 535

\bibitem[Binette et al.(1999)]{1999A&A...346..260B} Binette, L., Cabrit,
S., Raga, A., \& Cant\'{o}, J.\ 1999, \aap, 346, 260 %(B99)

\bibitem[Binette \& Robinson 1987]{bin87} Binette, L., \&
Robinson A. 1987, \aap, 177, 11

\bibitem[Binette et al.(2008)]{2008apj} Binette, L.,
Flores-Fajardo, N., S., Raga, A. C., Drissen, L., \& Morisset, C.
2008, \apj, in press (\pap)

%\bibitem[Bodo 1998]{bod98} Bodo, G.: 1998 in Astrophysical Jets:
%Open Problems, Massaglia S., Bodo G. (eds.), Amsterdam, Gordon and Breach, p.~161

\bibitem[Cant\'o \& Raga 1991]{can91} Cant\'o, J.,  \& Raga, A. C.
1991, \apj, 372, 646.

\bibitem[Cant\'{o} et al.(2000)]{2000ApJ...536..896C} Cant\'{o}, J., Raga,
A. C., \& Rodr\'{\i}guez, L.~F.\ 2000, \apj, 536, 896

\bibitem[Dionne \& Robert(2006)]{2006ApJ...641..252D} Dionne, D., \& Robert, C.\
2006, \apj, 641, 252

\bibitem[drissen00]{2000} Drissen, L., Roy, J.-R., Robert, C., Devost, D., \& Doyon, R. 2000,
\aj, 119, 688

\bibitem[drissen01]{2001} Drissen, L., Crowther, P. A., Smith, L. J., Robert, C., Roy, J.-R.,
\& Hillier, D. J. 2001, \apj, 546, 484

\bibitem[drissen09]{2009} Drissen, L., \'Ubeda, L., Charlebois, M.,
Binette, L., \& Roy, J.-R. 2009, \aj, in preparation (\dri)

%\bibitem[Dyson et~al. 1995]{dys95} Dyson, J.E., Hartquist, T.W.,
%Malone, M.T., Taylor, S.D., 1995, RMxAC 1, 119

\bibitem[Ferruit et al.(1997)]{1997A&A...322...73F} Ferruit, P., Binette,
L., Sutherland, R.~S., \& Pecontal, E.\ 1997, \aap, 322, 73

\bibitem[Gonzalez-Delgado et al.(1994)]{1994ApJ...437..239G}
Gonzalez-Delgado, R.~M., et al.\ 1994, \apj, 437, 239

\bibitem[Hartquist et al.(1986)]{1986MNRAS.221..715H} Hartquist, T.~W.,
Dyson, J.~E., Pettini, M., \& Smith, L.~J.\ 1986, \mnras, 221, 715

%\bibitem[Lavalley et~al. 1997]{lav97} Lavalley, C., Cabrit, S., Dougados,
%C., Ferruit, P.,  Bacon, R., 1997, A\&A 327, 671

\bibitem[Liedahl et al.(1995)]{1995ApJ...438L.115L} Liedahl, D.~A.,
Osterheld, A.~L., \& Goldstein, W.~H.\ 1995, \apjl, 438, L115

\bibitem[Luridiana et al.(1999)]{1999ApJ...527..110L} Luridiana, V.,
Peimbert, M., \& Leitherer, C.\ 1999, \apj, 527, 110

%\bibitem[Luridiana et al.(2001)]{2001A&A...379.1017L} Luridiana, V., Cervi\~{n}o, M., \&
%Binette, L.\ 2001, \aap, 379, 1017

%\bibitem[Madsen et al.(2006)]{2006ApJ...652..401M} Madsen, G.~J., Reynolds,
%R.~J., \& Haffner, L.~M.,\ 2006, \apj, 652, 401 (MRH)

%\bibitem[Martin(1997)]{1997ApJ...491..561M} Martin, C.~L.\ 1997, \apj, 491,
%561

%\bibitem[Melnick et al.(1999)]{1999MNRAS.302..677M} Melnick, J.,
%Tenorio-Tagle, G., \& Terlevich, R.\ 1999, \mnras, 302, 677

\bibitem[Melnick et al.(1988)]{1988MNRAS.235..297M} Melnick, J., Terlevich,
R., \& Moles, M.\ 1988, \mnras, 235, 297

\bibitem[Melnick et al.(2000)]{2000MNRAS.311..629M} Melnick, J., Terlevich,
R., \& Terlevich, E.\ 2000, \mnras, 311, 629

\bibitem[Mewe et al.(1986)]{1986A&AS...65..511M} Mewe, R., Lemen, J.~R., \& van den
Oord, G.~H.~J.\ 1986, \aaps, 65, 511

%\bibitem[Meaburn \& Dyson 1987]{dys87} Meaburn, J.,  Dyson, J.E.,
%1987, MNRAS 225, 863

\bibitem[Noriega-Crespo et al.(1996)]{1996ApJ...462..804N} Noriega-Crespo,
A., Garnavich, P.~M., Raga, A.~C., Canto, J., \& Boehm, K.-H.\ 1996,
\apj, 462, 804

%\bibitem[Otte et al.(2001)]{2001ApJ...560..207O} Otte, B., Reynolds, R.~J.,
%Gallagher, J.~S., III, \& Ferguson, A.~M.~N.\ 2001, \apj, 560, 207

%\bibitem[Otte et al.(2002)]{2002ApJ...572..823O} Otte, B., Gallagher,
%J.~S., III, \& Reynolds, R.~J.\ 2002, \apj, 572, 823

\bibitem[Pittard et al.(2005)]{2005MNRAS.361.1077P} Pittard, J.~M., Dyson,
J.~E., Falle, S.~A.~E.~G., \& Hartquist, T.~W.\ 2005, \mnras, 361,
1077

%\bibitem[Raga, Cabrit \& Cant\'o 1995]{rag95} Raga, A.C.,
%Cabrit, S., Cant\'o, J., 1995, MNRAS 273, 422

%\bibitem[Raga \& Cant\'o 1997]{rag97} Raga, A.C.,  Cant\'o, J.:
%1997, in Molecules in Astrophysics: Probes and Processes, Van
%Dishoeck E. (ed.), Dordrecht: Reidel, p.~89

%\bibitem[Raga, B\"ohm \& Cant\'o 1996]{rag96} Raga, A.C., B\"ohm, K.H.,
% Cant\'o, J., 1996, RMxAA 32, 161

%\bibitem[Rand(1996)]{1996ApJ...462..712R} Rand, R.~J.\ 1996, \apj, 462, 712

%\bibitem[Rand(1997)]{1997ApJ...474..129R} Rand, R.~J.\ 1997, \apj, 474, 129

\bibitem[Rand(1998)]{1998ApJ...501..137R} Rand, R.~J.\ 1998, \apj, 501,
137 (R98)

%\bibitem[Reynolds(1993)]{1993AIPC..278..156R} Reynolds, R.~J.\ 1993, Back
%to the Galaxy, 278, 156

%\bibitem[Reynolds \& Tufte(1995)]{1995ApJ...439L..17R} Reynolds, R.~J., \&
%Tufte, S.~L.\ 1995, \apjl, 439, L17

%\bibitem[Rossi et~al. 1997]{ros97} Rossi, P., Bodo, G., Massaglia, S.,
% Ferrari, A., 1997, A\&A 321, 672

%\bibitem[Raymond et~al. 1994]{ray94} Raymond, J.C., Morse, J.A.,
%Hartigan, P., Curiel, S.,  Heathcote, S., 1994, ApJ 434, 232

\bibitem[Roy et al.(1992)]{1992ApJ...386..498R} Roy, J.-R., Aube, M.,
McCall, M.~L., \& Dufour, R.~J.\ 1992, \apj, 386, 498

\bibitem[Slavin et al.(1993)]{1993ApJ...407...83S} Slavin, J.~D., Shull,
J.~M., \& Begelman, M.~C.\ 1993, \apj, 407, 83

%\bibitem[Solf 1987]{sol87} Solf, J., 1987, A\&A 180, 207

%\bibitem[Sternberg \& Soker(2008)]{2008MNRAS.384.1327S} Sternberg, A., \& Soker, N.\
%2008, \mnras, 384, 1327

\bibitem[Sternberg et al.(2003)]{2003ApJ...599.1333S} Sternberg, A.,
Hoffmann, T.~L., \& Pauldrach, A.~W.~A.\ 2003, \apj, 599, 1333

%\bibitem[Stone, Xu \& Hardee 1997]{sto97} Stone, J.M., Xu, J.,
% Hardee, P.E., 1997, ApJ 483, 136

%\bibitem[Taylor \& Raga 1995]{tay95} Taylor, S.D.,  Raga, A.C.
%1995, A\&A 296, 823

\bibitem[Tenorio-Tagle et al.(1997)]{1997ApJ...490L.179T} Tenorio-Tagle,
G., Munoz-Tunon, C., Perez, E., \& Melnick, J.\ 1997, \apjl, 490,
L179

\bibitem[Thuan \& Izotov(2005)]{2005ApJ...627..739T} Thuan, T.~X., \& Izotov,
Y.~I.\ 2005, \apj, 627, 739

\bibitem[Veilleux \& Osterbrock(1987)]{1987ApJS...63..295V} Veilleux, S., \&
Osterbrock, D.~E.\ 1987, \apjs, 63, 295

%\bibitem[Walterbos(1998)]{1998PASA...15...99W} Walterbos, R.~A.~M.\ 1998,
%Publications of the Astronomical Society of Australia, 15, 99

\bibitem[Westmoquette et al.(2007a)]{2007MNRAS.381..894W} Westmoquette,
M.~S., Exter, K.~M., Smith, L.~J., \& Gallagher, J.~S.\ 2007a,
\mnras, 381, 894

\bibitem[Westmoquette et al.(2007b)]{2007MNRAS.381..913W} Westmoquette,
M.~S., Smith, L.~J., Gallagher, J.~S., \& Exter, K.~M.\ 2007b,
\mnras, 381, 913

\bibitem[Westmoquette et al.(2007c)]{2007ApJ...671..358W} Westmoquette,
M.~S., Smith, L.~J., Gallagher, J.~S., III, O'Connell, R.~W.,
Rosario, D.~J., \& de Grijs, R.\ 2007c, \apj, 671, 358

\bibitem[Westmoquette et al.(2008)]{2008MNRAS.383..864W} Westmoquette,
M.~S., Smith, L.~J., \& Gallagher, J.~S.\ 2008, \mnras, 383, 864


\end{thebibliography}
\end{document}